# Derivative of the Lieb definition for the energy functional of density functional theory with respect to the particle number and the spin number

T. Gál* and P. Geerlings

General Chemistry Department (ALGC), Member of the QCMM Alliance Ghent-Brussels,
Free University of Brussels (VUB), 1050 Brussel, Belgium

**Abstract:** The nature of the explicit dependence on the particle number $N$ and on the spin number $N_s$ of the Lieb definition for the energy density functional is examined both in spin-independent and in spin-polarized density functional theory. It is pointed out that for ground states, the nonuniqueness of the external magnetic field $B(\vec{r})$ corresponding to a given pair of density $n(\vec{r})$ and spin density $n_s(\vec{r})$ in spin-polarized density functional theory requires the nonexistence of the derivative of the SDFT Lieb functional $F^L_{N,N_s}[n,n_s]$ with respect to $N_s$. Giving a suitable generalization of $F^L_N[n]$ and $F^L_{N,N_s}[n,n_s]$ for $N \neq \int n(\vec{r})d\vec{r}$ and $N_s \neq \int n_s(\vec{r})d\vec{r}$, it is then shown that their derivatives with respect to $N$ and $N_s$ are equal to the derivatives, with respect to $N$ and $N_s$, of the total energies $E[N,v]$ and $E[N,N_s,v,B]$ minus the external-field energy components, respectively.

*Email address: galt@phys.unideb.hu

## I. Introduction

The great success of the density functional theory (DFT) of many-electron systems [1,2] is due to the use of the electron density as basic variable in the place of the complicated many-variable, complex wavefunction. The cornerstone of DFT is the fact, discovered by Hohenberg and Kohn, that there exists a functional

$$E_v[n] = F[n] + \int n(\vec{r}) v(\vec{r}) d\vec{r} \tag{1}$$

of the electron density $n(\vec{r})$ whose minimum with respect to $n(\vec{r})$'s of a given norm $N$,

$$N = \int n(\vec{r}) d\vec{r} \;, \tag{2}$$

delivers the ground-state energy of an $N$-electron system in a given external potential $v(\vec{r})$, and the minimizing $n(\vec{r})$ is the ground-state density of the system. The universal functional $F[n]$ in Eq.(1) was originally defined only for $n(\vec{r})$'s that are ground-state densities for some external potential (i.e., are $v$-representable), which posed a substantial problem regarding the practical minimization of the energy functional $E_v[n]$. This problem was overcome by Levy's constrained-search definition for $F[n]$ [3,4],

$$F[n] = \min_{\psi_N \mapsto n} \langle \psi_N | \hat{T} + \hat{V}_{ee} | \psi_N \rangle \;, \tag{3}$$

where $\langle \psi | \hat{T} + \hat{V}_{ee} | \psi \rangle$ is minimized over the domain of normalized wavefunctions $\psi_N(\vec{r}_1 s_1, ..., \vec{r}_N s_N)$ that deliver a given $n(\vec{r})$ (which is denoted by $\psi_N \mapsto n$).

$F[n]$, as defined by Eq.(3), has some disadvantages. Most importantly, it is not a convex functional of the density. A functional $A[\rho]$ is said to be convex if

$$A[\alpha \rho_1 + (1-\alpha) \rho_2] \leq \alpha A[\rho_1] + (1-\alpha) A[\rho_2] \tag{4}$$

for $0 < \alpha < 1$. (For concavity, the inequality above is opposite.) Convexity is an essential element of mathematical analysis [5], and a convex $F[n]$ would have several favourable properties [2,4,6-8]. A convex $F[n]$ implies that the minimum-energy states are the only stationary points of $E_v[n]$ [4]. Further, convexity leads to differentiability quite naturally [6] (see also [2,7]). A physically important consequence of a convex $F[n]$ would also be a functionally size-consistent $E_v[n]$ [8].

Lieb [4] has given an alternative definition for the universal part of the energy density functional,

$$F_N^L[n] = \sup_v \left\{ E[N,v] - \int n(\vec{r}) v(\vec{r}) d\vec{r} \right\}, \tag{5}$$

where $E[N,v]$ denotes the ground-state energy of the $N$-electron system in external potential $v(\vec{r})$. This functional has an explicit dependence on $N = \int n(\vec{r}) d\vec{r}$ due to the term $E[N,v]$. To obtain the form to be inserted into Eq.(1), $\int n(\vec{r}) d\vec{r}$ has to be substituted for $N$ in Eq.(5),

$$F^L[n] = F_{\int n}^L[n] = \sup_v \left\{ E[\int n, v] - \int n(\vec{r}) v(\vec{r}) d\vec{r} \right\}. \tag{6}$$

The Lieb functional is obtained as the Legendre transform of (the minus of) the concave ground-state energy with respect to $v(\vec{r})$. This has the consequence that $F_N^L[n]$, and $F^L[n]$, is a convex functional. The inverse Legendre transformation gives back the Hohenberg-Kohn variational principle for $E_v[n] (= F_N^L[n] + \int n(\vec{r}) v(\vec{r}) d\vec{r})$,

$$E[N,v] = \inf_{n \mapsto N} \left\{ F_N^L[n] + \int n(\vec{r}) v(\vec{r}) d\vec{r} \right\}. \tag{7}$$

The Lieb functional can even be obtained via a constrained search construction [4], first proposed by Valone [9],

$$F^\Gamma[n] = \min_{\Gamma_N \mapsto n} \text{Tr}\left[ (\hat{T} + \hat{V}_{ee}) \Gamma_N \right], \tag{8}$$

where $\Gamma_N$ denotes the $N$-electron density matrix.

$F_N^L[n]$ has been generalized very recently by Ayers and Yang [10] (see also [11]) for spin-polarized DFT,

$$F_{N,N_s}^L[n,s] = \sup_{v,B} \left\{ E[N, N_s, v, B] - \int n(\vec{r}) v(\vec{r}) d\vec{r} + \int s(\vec{r}) \beta_e B(\vec{r}) d\vec{r} \right\}, \tag{9}$$

where an additional variable, the spin (polarization) density $s(\vec{r})$, appears due to the additional, magnetic external field $B(\vec{r})$. Eschrig [2] has generalized the Lieb functional for the spin-polarized case, too, but without an explicit dependence on the spin number; in this way, however, only ground states are treated. Further extensions of Lieb's Legendre transform idea have also been given [12,13].

In spite of the growing interest in the Lieb formulation of DFT [2,7,8,10,14-16] due to its mathematical advantages and its explicit treatment of particle number dependence, there is very little known about the actual nature of the dependence (of $F_N^L[n]$ and $F_{N,N_s}^L[n,s]$) on the particle number $N$, or on the spin number

$$N_s = \int s(\vec{r}) d\vec{r}. \tag{10}$$

One would intuitively expect some connection with the $N$ (or $N_s$) dependence of the energy $E[N,v]$ (or $E[N,N_s,v,B]$) itself, but because of the supremum with respect to the potentials in Eqs.(5) and (8), establishing such a relationship is a highly nontrivial task. One of the few known things is that $F_N^L[n]$ gives infinity for $n(\vec{r})$ with norm $\int n(\vec{r})d\vec{r}$ not equal to the $N$ value inserted into $F_N^L[n]$'s subscript [4] (see also [2]),

$$F_N^L[n] = +\infty \qquad \text{for} \quad \int n(\vec{r})d\vec{r} \neq N \ . \qquad (11)$$

The exploration of the nature of $F_N^L[n]$'s and $F_{N,N_s}^L[n,s]$'s dependences on their variables is of essential importance for the development of accurate density functionals, especially considering the renewed attention toward an explicit treatment of $N$ dependence [14,17,18]. It is worth underscoring that the Levy functional (Eq.(3)) also has an explicit particle number dependence [4], being equal to the Lieb functional for $v$-representable densities (see also [17]); however, it does not appear explicitly in its definition, just as in the case of the Valone definition for the Lieb functional, Eq.(8).

In this paper, the $N$ and $N_s$ dependence of $F_N^L[n]$ and $F_{N,N_s}^L[n,s]$ will be investigated. A suitable extension of $F_N^L[n]$ and $F_{N,N_s}^L[n,s]$ for $N \neq \int n(\vec{r})d\vec{r}$ and $N_s \neq \int s(\vec{r})d\vec{r}$ will be given. It will then be shown that their derivatives with respect to $N$ and $N_s$ are equal to the derivatives, with respect to $N$ and $N_s$, of the total energies $E[N,v]$ and $E[N,N_s,v,B]$ minus the external-field energy components, respectively. To say anything about derivatives with respect to $N$ and $N_s$, first, of course, a generalization of the functionals for fractional particle and spin numbers has to be given, which will be provided in Sec. II. In Sec. III, to illuminate the fact that there is physics behind the explicit $N$- and $N_s$-dependence of the Lieb functional, it will be shown that for ground states, the recently uncovered nonuniqueness of the external magnetic field $B(\vec{r})$ corresponding to a given pair of density $n(\vec{r})$ and spin density $s(\vec{r})$ [19,20] necessarily requires a discontinuity of the derivative of $F_{N,N_s}^L[n,s]$ with respect to $N_s$. The connection between the derivatives, with respect to $N$ and $N_s$, of $F_N^L[n]$ and $F_{N,N_s}^L[n,s]$ and of $E[N,v]$ and $E[N,N_s,v,B]$ will then be established in Sec. IV.

## II. Generalization of $F_N^L[n]$ and $F_{N,N_s}^L[n,s]$ for fractional particle numbers

To have a fractional particle number generalization for an energy density functional, first one should decide what meaning to be associated to the energy of, say, 4.3 electrons. That is, one should define $E[N,v]$ for fractional $N$'s. Physically, the best choice for a generalized $E[N,v]$ is the zero-temperature grand canonical ensemble definition [21,22]. That $E[N,v]$ can be given as

$$E[N,v] = \inf_{\hat{\Gamma} \mapsto N} \text{Tr}\left[\hat{H}_v \hat{\Gamma}\right], \tag{12}$$

where the infimum is searched over statistical mixtures $\hat{\Gamma}$ that give particle number $N$, $\text{Tr}[\hat{N}\hat{\Gamma}] = N$. Provided the ground-state energy $E[M,v]$ of systems of integer number of electrons is a convex function of the electron number at fixed $v(\vec{r})$ (for which there is experimental, and also numerical, evidence [21]), the above definition yields the energy for a general particle number as

$$E[N,v] = (1-\omega)E[M,v] + \omega E[M+1,v], \tag{13}$$

where $M$ is the integer part of $N$, and $\omega$ is the fractional part of $N$ (i.e., $\omega = N - M$). Having an extension of $E[N,v]$ concave in $v(\vec{r})$, the Lieb functional can be easily generalized for fractional particle numbers by inserting the extended $E[N,v]$ into Eq.(5). The generalization obtained with the use of Eq.(12) has been given by Eschrig [2]. For this generalized $F_N^L[n]$, the following important property holds:

$$\begin{aligned} F_N^L[(1-\omega)n_M + \omega n_{M+1}] \\ &= \sup_v \left\{ (1-\omega)E[M,v] + \omega E[M+1,v] - \int \left((1-\omega)n_M(\vec{r}) + \omega n_{M+1}(\vec{r})\right)v(\vec{r})d\vec{r} \right\} \\ &= (1-\omega)\sup_v \left\{ E[M,v] - \int n_M(\vec{r})v(\vec{r})d\vec{r} \right\} + \omega \sup_v \left\{ E[M+1,v] - \int n_{M+1}(\vec{r})v(\vec{r})d\vec{r} \right\} \\ &= (1-\omega)F_M^L[n_M] + \omega F_{M+1}^L[n_{M+1}], \end{aligned} \tag{14}$$

where $n_M(\vec{r})$ and $n_{M+1}(\vec{r})$ are $M$-electron and $(M+1)$-electron densities, respectively, in the same external potential.

The fractional particle number generalization of $E[N,N_s,v,B]$, too, can be based on the zero-temperature grand canonical ensemble. It leads to a definition

$$E[N,N_s,v,B] = \inf_{\hat{\Gamma} \mapsto N,N_s} \text{Tr}\left[\hat{H}_{v,B} \hat{\Gamma}\right], \tag{15}$$

where the states of which a statistical mixture $\hat{\Gamma} = \sum_j |\Psi_j\rangle p_j \langle\Psi_j|$ is composed are not required to have the given $N_s$ separately, but only their averaged spin numbers have to give $N_s$. In obtaining a spin-polarized version of Eq.(13), a subtle point is that since SDFT treats the lowest-energy states of *every* spin multiplicity, there are many *M*-electron and (*M*+1)-electron states, which have to be "paired" in some way to obtain proper weighted averages corresponding to the (*M*+ω)-electron states. Yang and coworkers have recently given some insight into the necessary shape of $E(N, N_s)$ [23], regardless of the concrete definition of $E[N, N_s, v, B]$ for fractional *N* and $N_s$, relying on their infinite separation approach [22], but only in the case of ground states without an external magnetic field.

Accepting Eq.(15) as the fractional particle number extension of the energy, and inserting it into Eq.(9), a generalization of the SDFT Lieb functional for fractional *N* is obtained. Since the Lieb functional is constructed via the Legendre transformation of (minus) the energy, a consequence of the concavity of the energy Eq.(15) in $(v(\vec{r}), B(\vec{r}))$ is that it can be obtained by

$$E[N, N_s, v, B] = \inf_{n \mapsto N, s \mapsto N_s} \left\{ F^L_{N,N_s}[n, s] + \int n(\vec{r}) v(\vec{r}) d\vec{r} - \int s(\vec{r}) \beta_e B(\vec{r}) d\vec{r} \right\}. \tag{16}$$

Eq.(16) is the corresponding generalization of the Hohenberg-Kohn variational principle. That Eq.(15) is concave in $(v(\vec{r}), B(\vec{r}))$ can be proved in the following way:

$$E[N, N_s, \alpha v_1 + (1-\alpha)v_2, \alpha B_1 + (1-\alpha)B_2] = \inf_{\hat{\Gamma} \mapsto N, N_s} \left\{ \alpha \operatorname{Tr}[\hat{H}_{v_1, B_1} \hat{\Gamma}] + (1-\alpha) \operatorname{Tr}[\hat{H}_{v_2, B_2} \hat{\Gamma}] \right\}$$

$$\geq \alpha \inf_{\hat{\Gamma} \mapsto N, N_s} \operatorname{Tr}[\hat{H}_{v_1, B_1} \hat{\Gamma}] + (1-\alpha) \inf_{\hat{\Gamma} \mapsto N, N_s} \operatorname{Tr}[\hat{H}_{v_2, B_2} \hat{\Gamma}] = \alpha E[N, N_s, v_1, B_1] + (1-\alpha) E[N, N_s, v_2, B_2] , \tag{17}$$

where the fact that the infimum of the sum of two terms cannot be lower than the sum of the independent infima of the terms is utilized. The equality holds only in cases where $B_1(\vec{r})$ and $B_2(\vec{r})$ correspond to the same lowest-lying energy-eigenstate with $N_s$. (The ambiguity of $v(\vec{r})$ is fixed, to define the zero of the energy.)

It is worth giving the generalization of the Lieb functional in the $(N_\uparrow, N_\downarrow)$ representation, too, where the spin-up and spin-down densities,

$$n_\uparrow(\vec{r}) = \frac{1}{2}(n(\vec{r}) + s(\vec{r})) \tag{18a}$$

and

$$n_\downarrow(\vec{r}) = \frac{1}{2}(n(\vec{r}) - s(\vec{r})), \tag{18b}$$

are the basic variables. The energy $E[N_\uparrow, N_\downarrow, v, B]$ for fractional $N$ can be deduced from Eq.(15) via the use of the transformation Eq.(18), that is,

$$E[N_\uparrow, N_\downarrow, v, B] = E[N = N_\uparrow + N_\downarrow, N_s = N_\uparrow - N_\downarrow, v, B] \ . \tag{19}$$

Inserting Eq.(19), with Eq.(15), into

$$F^L_{N_\uparrow, N_\downarrow}[n_\uparrow, n_\downarrow] = \sup_{v,B} \left\{ E[N_\uparrow, N_\downarrow, v, B] - \int n_\uparrow(\vec{r})\big(v(\vec{r}) - \beta_e B(\vec{r})\big)d\vec{r} - \int n_\downarrow(\vec{r})\big(v(\vec{r}) + \beta_e B(\vec{r})\big)d\vec{r} \right\}, \tag{20}$$

the desired generalization is obtained. Eschrig [2] has also given the fractional particle number generalization of a spin-polarized Lieb functional; however, treating only ground states.

### III. Euler-Lagrange equations for the Lieb functionals, and the effect of $B(\vec{r})$'s nonuniqueness on $F^L_{N,N_s}[n,s]$'s derivative with respect to $N_s$

#### Euler-Lagrange equations in the spin-independent case

The Euler-Lagrange equation emerging from the variational principle for $E^L_{N,v}[n] = F^L_N[n] + \int n(\vec{r})v(\vec{r})d\vec{r}$ for the determination of the ground-state density corresponding to a given $v(\vec{r})$ and $N$ is

$$\left.\frac{\delta F^L_N[n]}{\delta n(\vec{r})}\right|_N + v(\vec{r}) = \mu_N \ . \tag{21}$$

In Eq.(21), the derivative of $F^L_N[n]$ has to be restricted to the domain of $n(\vec{r})$'s with the given $N$, since $F^L_N[n]$ gives infinity for $\int n(\vec{r})d\vec{r} \neq N$ [4] (see also [2]), therefore its full derivative with respect to $n(\vec{r})$ does not exist. That is, the derivative in Eq.(21) is an $N$-restricted derivative, determined only up to an additive constant (with respect to $\vec{r}$) [for a discussion of restricted derivatives, see Sec.II of [24]]. This means that $\mu_N$ is ambiguous, too.

However, another Euler-Lagrange equation can be obtained if instead of $E^L_{N,v}[n]$, $E^L_v[n] = F^L[n] + \int n(\vec{r})v(\vec{r})d\vec{r}$ is minimized to determine the ground-state density; namely,

$$\frac{\delta F^L[n]}{\delta n(\vec{r})} + v(\vec{r}) = \mu \ . \tag{22}$$

In the above equation, the derivative does not have to be restricted, since in the minimization of $E_v^L[n] = E_{\int n, v}^L[n]$, $N$ varies together with $n(\vec{r})$. The Lagrange multiplier $\mu$ emerging from the conservation constraint of the particle number (Eq.(2)) in the minimization can be identified with the derivative of the ground-state energy $E[N,v]$ with respect to the particle number $N$, just as in the case of the constrained search definition for $F[n]$. That is,

$$\mu = \frac{\partial E[N,v]}{\partial N} \quad . \tag{23}$$

Note that the derivatives in Eqs.(22) and (23) will be one-sided at integer particle numbers because of the derivative discontinuities there.

Utilizing that $F^L[n] = F_N^L[n]\big|_{N=\int n}$, Eq.(22) formally gives

$$\frac{\delta F_N^L[n]}{\delta n(\vec{r})} + v(\vec{r}) + \frac{\partial F_N^L[n]}{\partial N} = \mu \quad . \tag{24}$$

$\mu$, thus, is connected to $\mu_N$ by

$$\mu_N = \mu - \frac{\partial F_N^L[n]}{\partial N} \quad . \tag{25}$$

However, $F_N^L[n]$'s definition gives infinity for $n(\vec{r})$'s with $\int n(\vec{r})d\vec{r} \neq N$, that is, $F_N^L[n]$'s values for the domain of $n(\vec{r})$'s of $\int n(\vec{r})d\vec{r} = N$ are in a valley with infinitely high walls. This has the consequence that $\frac{\partial F_N^L[n]}{\partial N}$ does not exist (since the derivative with respect to $N$ is taken at fixed $n(\vec{r})$, going out of the $N = \int n(\vec{r})d\vec{r}$ domain), and $F_N^L[n]$ may have only a restricted derivative $\frac{\delta F_N^L[n]}{\delta n(\vec{r})}\bigg|_N$ with respect to $n(\vec{r})$ (for $n(\vec{r})$'s of $\int n(\vec{r})d\vec{r} = N$), as already noted above. That $F_N^L[n]$ actually *has* a derivative (with respect to $n(\vec{r})$) for v-representable densities over the domain $\int n(\vec{r})d\vec{r} = N$ has been proven recently by Lammert [25], revising the earlier proof by Englisch and Englisch [6], built on the convexity of $F_N^L[n]$.

To have finite values also for $n(\vec{r})$'s of $\int n(\vec{r})d\vec{r} \neq N$, $F_N^L[n]$ can be modified as

$$\tilde{F}_N^L[n] = \left(\frac{\int n}{N}\right) F_N^L\left[N \frac{n}{\int n}\right] , \tag{26}$$

e.g. This kind of modification of $F_N^L[n]$ to eliminate the infinite values has been proposed by Lieb himself [4]; however, in his Eq.(3.18), the $N$ and $1/N$ factors are missing, giving an inappropriate formula. If $F_N^L[n]$ is differentiable over the domain $\int n(\vec{r})d\vec{r} = N$, then $\tilde{F}_N^L[n]$ has a full derivative, since [26] (i) $N\dfrac{n(\vec{r})}{\int n(\vec{r}')d\vec{r}'}$ is fully differentiable, and (ii) it integrates to $N$ for any $n(\vec{r})$ (plus of course $\dfrac{\int n(\vec{r})d\vec{r}}{N}$ is differentiable as well). Note that instead of the above, degree-one homogeneous extension of $F_N^L[n]$ from the domain $\int n(\vec{r})d\vec{r} = N$, other extensions could be applied as well; see Eq.(8) in [27], with $g(\vec{r}) = 1$ and $L=N$, e.g. The simplest extension would be the constant shifting of $F_N^L[n]$ (cancelling the factor $\dfrac{\int n}{N}$ in Eq.(26)), that is, the degree-zero homogeneous extension. It is worth mentioning, however, that the degree-one homogeneous extension is the one that is in accordance with the structure of Schrödinger quantum mechanics [17], on the basis of which it has been proposed that the density functionals have a degree-one homogeneous density dependence, beside an explicit $N$-dependence [17].

With the modified $F_N^L[n]$, Eqs.(24) and (25) can be correctly written. For different modifications $\tilde{F}_N^L[n]$, the derivative $\dfrac{\partial \tilde{F}_N^L[n]}{\partial N}$, and the Lagrange multiplier $\mu_N$, will of course be different. Note however that $\mu$ will be the same for every $\tilde{F}_N^L[n]$, since $\tilde{F}_{\int n}^L[n] = F_{\int n}^L[n]$.

**Euler-Lagrange equations in the spin-polarized case**

Similar to the spin-free case, $F_{N,N_s}^L[n,s]$ can be modified for $n(\vec{r})$'s of $\int n(\vec{r})d\vec{r} \neq N$, and for $s(\vec{r})$'s of $\int s(\vec{r})d\vec{r} \neq N_s$, to have well-defined values everywhere, and to be fully differentiable with respect to $(n(\vec{r}), s(\vec{r}))$ (assuming that Lammert's proof can be generalized for the spin-polarized case). With this differentiable extension (not required to be the degree-one homogeneous extension), denoted by $\tilde{F}_{N,N_s}^L[n,s]$, the Euler-Lagrange equations arising from the variational principle for $E_{N,N_s,v,B}^L[n,s]$ for the determination of the density of the lowest-energy state with $(N, N_s)$ in a given $(v(\vec{r}), B(\vec{r}))$ can be written as

$$\frac{\delta \widetilde{F}^L_{N,N_s}[n,s]}{\delta n(\vec{r})} + v(\vec{r}) = \mu_N \tag{27}$$

and

$$\frac{\delta \widetilde{F}^L_{N,N_s}[n,s]}{\delta s(\vec{r})} - \beta_e B(\vec{r}) = \mu_{N_s} \quad . \tag{28}$$

The corresponding Euler-Lagrange equations that emerge from the minimization of $E^L_{v,B}[n,s]$ instead of $E^L_{N,N_s,v,B}[n,s]$ are

$$\frac{\delta \widetilde{F}^L_{N,N_s}[n,s]}{\delta n(\vec{r})} + v(\vec{r}) + \frac{\partial \widetilde{F}^L_{N,N_s}[n,s]}{\partial N} = \mu \tag{29}$$

and

$$\frac{\delta \widetilde{F}^L_{N,N_s}[n,s]}{\delta s(\vec{r})} - \beta_e B(\vec{r}) + \frac{\partial \widetilde{F}^L_{N,N_s}[n,s]}{\partial N_s} = \mu_s \quad . \tag{30}$$

The connection between the Lagrange multipliers of the two pairs of Euler-Lagrange equations is given by

$$\mu_N = \mu - \frac{\partial \widetilde{F}^L_{N,N_s}[n,s]}{\partial N} \tag{31}$$

and

$$\mu_{N_s} = \mu_s - \frac{\partial \widetilde{F}^L_{N,N_s}[n,s]}{\partial N_s} \quad . \tag{32}$$

The Lagrange multipliers $\mu$ and $\mu_s$ (not $\mu_N$ and $\mu_{N_s}$ !) can be identified as the derivatives of the energy $E[N, N_s, v, B]$ with respect to $N$ and $N_s$, respectively, similar to the case of the constrained search definition for $F[n,s]$ [28]. That is,

$$\mu = \frac{\partial E[N, N_s, v, B]}{\partial N} \tag{33}$$

and

$$\mu_s = \frac{\partial E[N, N_s, v, B]}{\partial N_s} \quad . \tag{34}$$

**Effect of $B(\vec{r})$'s nonuniqueness**

Eschrig and Pickett [19] and Capelle and Vignale [20] have shown recently that the correspondence between $(n(\vec{r}), s(\vec{r}))$ and $(v(\vec{r}), B(\vec{r}))$ is not one-to-one for nondegenerate ground states, but $B(\vec{r})$ is determined by $(n(\vec{r}), s(\vec{r}))$ only up to a nontrivial additive constant [19] (see also [10,29-31]). This nonuniqueness of the external magnetic field $B(\vec{r})$ implies for ground states the nonexistence of the full derivative of the energy density functional $E_{v,B}[n,s]$ with respect to $s(\vec{r})$. Fortunatelly, $B(\vec{r})$'s nonuniqueness does not also exclude the existence of one-sided derivatives with respect to $s(\vec{r})$ [28], which means that there is only a simple derivative discontinuity at the given $s(\vec{r})$'s with integer norm $N_s$. The question naturally arises: what are the implications of $B(\vec{r})$'s nonuniqueness for the Lieb energy functional $E^L_{N,N_s,v,B}[n,s]$, which has an explicit dependence on $N_s$?

A ground state *can always* be obtained from $E^L_{N,N_s,v,B}[n,s]$ by minimizing it under the constraint of conserving only $N = \int n(\vec{r})d\vec{r}$. Therefore the following Euler-Lagrange equations arise for the ground-state $(n(\vec{r}), s(\vec{r}))$ if $F^L_{N,N_s}[n,s]$ has the corresponding derivatives with respect to $n(\vec{r})$, $s(\vec{r})$, and $N$ and $N_s$:

$$\frac{\delta \tilde{F}^L_{N,N_s}[n,s]}{\delta n(\vec{r})} + v(\vec{r}) + \frac{\partial \tilde{F}^L_{N,N_s}[n,s]}{\partial N} = \mu \qquad (35)$$

and

$$\frac{\delta \tilde{F}^L_{N,N_s}[n,s]}{\delta s(\vec{r})} - \beta_e B(\vec{r}) + \frac{\partial \tilde{F}^L_{N,N_s}[n,s]}{\partial N_s} = 0 \; . \qquad (36)$$

It can be seen that Eq.(36) leads to a contradiction due to $B(\vec{r})$'s ambiguity, since it has to hold also for a $B(\vec{r}) + \Delta B$, because of the fact that the same ground state $(n(\vec{r}), s(\vec{r}))$ can be obtained from magnetic fields differing by a constant. This indicates that $\dfrac{\partial \tilde{F}^L_{N,N_s}[n,s]}{\partial N_s}$ does not exist. Consequently, either there is a derivative discontinuity in $\dfrac{\partial \tilde{F}^L_{N,N_s}[n,s]}{\partial N_s}$ ($n(\vec{r})$ and $s(\vec{r})$ fixed), or even the one-sided derivatives of $F^L_{N,N_s}[n,s]$ with respect to $N_s$ do not exist. This is true for any modification of $F^L_{N,N_s}[n,s]$ for $s(\vec{r})$'s of $\int s(\vec{r})d\vec{r} \neq N_s$, that is, $F^L_{N,N_s}[n,s]$ cannot be differentiated with respect to its $N_s$ dependence.

It has to be noted that another resolution of the contradiction caused by $B(\vec{r})$'s ambiguity in Eq.(36) could be that $F^L_{N,N_s}[n,s]$ does not have derivative with respect to $s(\vec{r})$ over the domain $\int s(\vec{r})d\vec{r} = N_s$, i.e., the proof of $F^L_N[n]$'s differentiability with respect to the density cannot be extended to the spin-polarized case. This would of course imply quite sad consequences for SDFT, the determination of ground states via Euler-Lagrange equations becoming impossible. Note though that the generally applied, Kohn-Sham, formulation of DFT can be established also without the use of functional derivatives [32].

## IV. The derivatives of $F^L_N[n]$ and $F^L_{N,N_s}[n,s]$ with respect to $N$ and $N_s$

As can be seen from their definitions, the explicit $N$- and $N_s$-dependence of $F^L_N[n]$ and $F^L_{N,N_s}[n,s]$ are determined by the $N$- and $N_s$-dependence of the energy itself. However, these connections are highly nontrivial because of the supremums with respect to $v(\vec{r})$ and $B(\vec{r})$. (For example, differentiating $\sup_v\{f[N,v]\}$ with respect to $N$ does not equal $\sup_v\left\{\dfrac{\partial f[N,v]}{\partial N}\right\}$ generally.) Further, their actual form is affected by the chosen modifications of the original $F^L_N[n]$ and $F^L_{N,N_s}[n,s]$ to have finite values for densities with norms differing from the ones given in the subscripts. It will be shown here that by choosing $\tilde{F}^L_N[n]$ and $\tilde{F}^L_{N,N_s}[n,s]$ properly, their derivatives with respect to $N$ and $N_s$ turn out to have a very natural relationship with the corresponding derivatives of the energy.

### A. The spin-independent case

To define an $\tilde{F}^L_N[n]$ for $n(\vec{r})$'s with $\int n(\vec{r})d\vec{r} \neq N$, a mapping $n_N[n]$ from $n(\vec{r})$'s of arbitrary norm onto $n_N(\vec{r})$'s of norm $N$ has to be given, with which then $\tilde{F}^L_N[n] = F^L_N[n_N[n]]$. (In Eq.(26), $n_N[n] = N\dfrac{n}{\int n}$; the $\dfrac{\int n}{N}$ factor before $F^L_N$ is irrelevant with this respect.) In the zero-temperature grand canonical ensemble generalization, the density of an arbitrary norm $N$ emerges as

$$n(\vec{r}) = (1-\omega) n_M(\vec{r}) + \omega n_{M+1}(\vec{r}) , \qquad (37)$$

where $n(\vec{r})$, $n_M(\vec{r})$ and $n_{M+1}(\vec{r})$ correspond to the same external potential $v(\vec{r})$ ($n(\vec{r})$ determines $v(\vec{r})$, hence $n_M(\vec{r})$ and $n_{M+1}(\vec{r})$, uniquely [33]). To obtain the proper $\tilde{F}_N^L[n]$, we define the necessary $n(\vec{r}) \to n_N(\vec{r})$ mapping in the following way: We associate a $n(\vec{r})$ of $\int n(\vec{r})d\vec{r} \neq N$ with the $n_N(\vec{r})$ that corresponds to the same external potential. For non-$v$-representable $n(\vec{r})$'s, we utilize the fact that the ensemble-$v$-representable densities are dense in the set of all ($N$-representable) $n(\vec{r})$'s [6], that is, for any non-$v$-representable $n(\vec{r})$ there is a sequence of ensemble-$v$-representable densities $n^{(i)}(\vec{r})$ that converges to the given $n(\vec{r})$. We then define $n_N[n]$ for non-$v$-representable $n(\vec{r})$ by $\lim_i n_N[n^{(i)}]$. (This is similar to how Ayers gives an alternative definition for $F_N^L[n]$ in Ref.[8].)

With the above choice, $\tilde{F}_N^L[n]$'s derivative with respect to $N$ for a given (ensemble-) $v$-representable $n(\vec{r})$ with $\int n(\vec{r})d\vec{r} = N$ can be calculated as

$$\left.\frac{\partial \tilde{F}_N^L[n]}{\partial N}\right|_+ = \lim_{\varepsilon \to 0+} \frac{F_{N+\varepsilon}^L[n_{N+\varepsilon}[n]] - F_N^L[n]}{\varepsilon}$$

$$= \lim_{\varepsilon \to 0+} \frac{\sup_v \{E[N+\varepsilon,v] - \int n_{N+\varepsilon}(\vec{r})v(\vec{r})d\vec{r}\} - \sup_v \{E[N,v] - \int n(\vec{r})v(\vec{r})d\vec{r}\}}{\varepsilon} . \qquad (38)$$

(One-sided derivative is calculated because of the possible discontinuity.) Since for $v$-representable densities, the supremum in $F_N^L[n]$'s definition is achieved at the $v(\vec{r})$ the density in $F_N^L[n]$'s argument corresponds to, and $n_{N+\varepsilon}(\vec{r})$ belongs to the same $v(\vec{r})$ for any $\varepsilon$, Eq.(38) can be written as

$$\left.\frac{\partial \tilde{F}_N^L[n]}{\partial N}\right|_+ = \lim_{\varepsilon \to 0+} \frac{\left(E[N+\varepsilon,v] - \int n(\vec{r})[N+\varepsilon,v]\,v(\vec{r})d\vec{r}\right) - \left(E[N,v] - \int n(\vec{r})[N,v]v(\vec{r})d\vec{r}\right)}{\varepsilon} . \qquad (39)$$

Eq.(39) finally gives

$$\left.\frac{\partial \tilde{F}_N^L[n]}{\partial N}\right|_+ = \left.\frac{\partial \left(E[N,v] - \int n(\vec{r})[N,v]\,v(\vec{r})d\vec{r}\right)}{\partial N}\right|_+ . \qquad (40)$$

Of course, a similar derivation applies for the left-side derivative as well; thus, Eq.(40) can be written also with a minus instead of the plus in the subscripts.

The above formula is a result that on one hand might be expected on the basis of $F_N^L[n]$'s definition, but at the same time can be quite suprising if one considers that on the left of Eq.(40), $N$ is varied with the density being fixed, while on the right, $N$ is varied with the external potential being fixed. It is important to recognize that the density dependence of the left side of Eq.(40) does not disappear on the right side; it is present (though not denoted for simplicity) through $v(\bar{r})[n]$, as can be seen from the derivation. Since the energy derivative with respect to $N$ is just the chemical potential, and the density derivative with respect to $N$ is the Fukui function [34], Eq.(40) can also be written as

$$\left.\frac{\partial \tilde{F}_N^L[n]}{\partial N}\right|_+ = \mu^+ - \int f^+(\bar{r}) v(\bar{r}) \, d\bar{r} \ . \tag{41}$$

With the use of Eq.(40), $\mu_N$ of Eq.(21) (with $\tilde{F}_N^L[n]$ in the place of $F_N^L[n]$) can be given as well,

$$\mu_N^+ = \int f^+(\bar{r}) v(\bar{r}) \, d\bar{r} \ , \tag{42}$$

utilizing Eq.(33). Note that without a modification of $F_N^L[n]$, Eq.(21) could be written only with the ambiguous restricted derivative $\left.\frac{\delta F_N^L[n]}{\delta n(\bar{r})}\right|_N$, and with an ambiguous $\mu_N$. It is also worth mentioning that with the modification Eq.(26), the derivative with respect to $N$ can also be calculated, utilizing the degree-one homogeneity in $n(\bar{r})$ of that expression in Eq.(24); namely, $\left.\partial \tilde{F}_N^L[n]/\partial N\right|_+ = \mu^+ - E/N$. (However, in this case, the result is not an expression calculated on the basis of definition.)

### B. The spin-polarized generalization

For the spin-polarized version of the Lieb functional, an expression analogous to Eq.(40) can be derived both for the $N$- and for the $N_s$-dependence. We now map a pair of $n(\bar{r})$ and $s(\bar{r})$ of arbitrary norms, corresponding to a state with external fields $v(\bar{r})$ and $B(\bar{r})$, onto a pair of $n_N(\bar{r})$ and $s_{N_s}(\bar{r})$ of norms $N$ and $N_s$ that corresponds to the same $v(\bar{r})$ and $B(\bar{r})$. Because of $B(\bar{r})$'s nonuniqueness, however, we have to choose among the possible $B(\bar{r})$'s [$B(\bar{r}) + \Delta B$, with $0 \leq \Delta B \leq \Delta B_{max}$] that yield the same $n(\bar{r})$ and $s(\bar{r})$: we choose the one that is halfway between two energy-level crossings, i.e., that corresponds to $\Delta B_{max}/2$. With this mapping, $\tilde{F}^L_{N,N_s}[n,s] = F^L_{N,N_s}[(n_N, s_{N_s})[n,s]]$.

$\tilde{F}^L_{N,N_s}[n,s]$'s derivative with respect to $N$ for a given $(v,B)$-representable $n(\vec{r})$ and $s(\vec{r})$ with $\int n(\vec{r})d\vec{r} = N$ and $\int s(\vec{r})d\vec{r} = N_s$ can be calculated as

$$\left.\frac{\partial \tilde{F}^L_{N,N_s}[n,s]}{\partial N}\right|_+ = \lim_{\varepsilon \to 0+} \frac{F^L_{N+\varepsilon,N_s}[(n_{N+\varepsilon}, s_{N_s})[n,s]] - F^L_{N,N_s}[n,s]}{\varepsilon} \ . \tag{43}$$

Because of similar arguments as in the spin-independent case, we obtain

$$\left.\frac{\partial \tilde{F}^L_{N,N_s}[n,s]}{\partial N}\right|_+ = \lim_{\varepsilon \to 0+} \frac{1}{\varepsilon}\left\{\left(E[N+\varepsilon, N_s, v, B] - \int n(\vec{r})[N+\varepsilon, N_s, v, B]\, v(\vec{r})\, d\vec{r} + \int s(\vec{r})[N+\varepsilon, N_s, v, B]\, \beta_e B(\vec{r})\, d\vec{r}\right) \right.$$
$$\left. - \left(E[N, N_s, v, B] - \int n(\vec{r})[N, N_s, v, B]\, v(\vec{r})\, d\vec{r} + \int s(\vec{r})[N, N_s, v, B]\, \beta_e B(\vec{r})\, d\vec{r}\right)\right\}, \tag{44}$$

which then gives

$$\left.\frac{\partial \tilde{F}^L_{N,N_s}[n,s]}{\partial N}\right|_+ = \left.\frac{\partial\left(E[N, N_s, v, B] - \int n(\vec{r})[N, N_s, v, B]\, v(\vec{r})d\vec{r} + \int s(\vec{r})[N, N_s, v, B]\, \beta_e B(\vec{r})d\vec{r}\right)}{\partial N}\right|_+ . \tag{45}$$

$\tilde{F}^L_{N,N_s}[n,s]$'s derivative with respect to $N_s$ for a given $(v,B)$-representable $n(\vec{r})$ and $s(\vec{r})$ with $\int n(\vec{r})d\vec{r} = N$ and $\int s(\vec{r})d\vec{r} = N_s$ can be calculated analogously to the derivative with respect to $N$. That is,

$$\left.\frac{\partial \tilde{F}^L_{N,N_s}[n,s]}{\partial N_s}\right|_+ = \lim_{\varepsilon \to 0+} \frac{F^L_{N,N_s+\varepsilon}[(n_N, s_{N_s+\varepsilon})[n,s]] - F^L_{N,N_s}[n,s]}{\varepsilon} \ . \tag{46}$$

Since again, for $(v,B)$-representable $n(\vec{r})$ and $s(\vec{r})$, the supremum in $F^L_{N,N_s}[n,s]$'s definition is achieved at the $v(\vec{r})$ and $B(\vec{r})$ the density and spin density in $F^L_{N,N_s}[n,s]$'s argument correspond to, Eq.(46) gives

$$\left.\frac{\partial \tilde{F}^L_{N,N_s}[n,s]}{\partial N_s}\right|_+ = \lim_{\varepsilon \to 0+} \frac{1}{\varepsilon}\left\{\left(E[N, N_s+\varepsilon, v, B] - \int n(\vec{r})[N, N_s+\varepsilon, v, B]\, v(\vec{r})d\vec{r} + \int s(\vec{r})[N, N_s+\varepsilon, v, B]\, \beta_e B(\vec{r})\, d\vec{r}\right)\right.$$
$$\left. - \left(E[N, N_s, v, B] - \int n(\vec{r})[N, N_s, v, B]\, v(\vec{r})\, d\vec{r} + \int s(\vec{r})[N, N_s, v, B]\, \beta_e B(\vec{r})\, d\vec{r}\right)\right\}. \tag{47}$$

Eq.(47) then yields

$$\left.\frac{\partial \tilde{F}^L_{N,N_s}[n,s]}{\partial N_s}\right|_+ = \left.\frac{\partial\left(E[N, N_s, v, B] - \int n(\vec{r})[N, N_s, v, B]\, v(\vec{r})\, d\vec{r} + \int s(\vec{r})[N, N_s, v, B]\, \beta_e B(\vec{r})d\vec{r}\right)}{\partial N_s}\right|_+ . \tag{48}$$

Eq.(48) and Eq.(45) are of course valid with left-side derivatives, too. They can also be written with the use of the chemical potential, the spin chemical potential, and the generalized Fukui functions [35], as

$$\left.\frac{\partial \widetilde{F}_{N,N_s}^L[n,s]}{\partial N}\right|_+ = \mu^+ - \int f_{NN}^+(\vec{r})v(\vec{r})\,d\vec{r} + \int f_{SN}^+(\vec{r})\beta_e B(\vec{r})d\vec{r} \quad (49)$$

and

$$\left.\frac{\partial \widetilde{F}_{N,N_s}^L[n,s]}{\partial N_s}\right|_+ = \mu_s^+ - \int f_{NS}^+(\vec{r})v(\vec{r})\,d\vec{r} + \int f_{SS}^+(\vec{r})\beta_e B(\vec{r})d\vec{r} \quad . \quad (50)$$

**V. Summary**

We studied the $N$- and $N_s$-dependence of the spin-free, $F_N^L[n]$, and the spin-polarized version, $F_{N,N_s}^L[n,s]$, of the Lieb functional of density functional theory. To investigate those dependences analytically, a modification of the Lieb functionals' definitions is necessary, since the original definitions give infinity for densities with norm not equal to that given in their subscripts. Since $F_N^L[n]$ and $F_{N,N_s}^L[n,s]$ have physical relevance only for $n(\vec{r})$ with $\int n(\vec{r})d\vec{r} = N$ and for $s(\vec{r})$ with $\int s(\vec{r})d\vec{r} = N_s$, that modification can be done freely. Of course, among the possibilities, that one is worth choosing that has physics behind it. This is similar to the fractional particle number generalization of the energy density functional, where the zero-temperature grand canonical ensemble extension is chosen, which gives a Lagrange multiplier in the minimization of the energy functional that equals the derivative of the energy with respect to the particle number. We have shown that with suitable extensions for $\int n(\vec{r})d\vec{r} \neq N$ and $\int s(\vec{r})d\vec{r} \neq N_s$, the Lieb functionals' derivatives with respect to the particle number and the spin number are equal to the derivatives with respect to $N$ and $N_s$, of the total energies $E[N,v]$ and $E[N,N_s,v,B]$ minus the external-field energy components, respectively, for ensemble-$v$, or ensemble-$(v,B)$, -representable densities and spin densities. The fractional particle number and spin number generalization of the Lieb functionals, which is necessary if one wants to differentiate with respect to $N$ and $N_s$, is given in Sec.II. In Sec.III, we have also shown how the nonuniqueness of the external magnetic field requires a discontinuity in the derivative of $F_{N,N_s}^L[n,s]$ with respect to $N_s$ (irrespective of $F_{N,N_s}^L[n,s]$'s modification for $\int s(\vec{r})d\vec{r} \neq N_s$), which in Sec.IV turns out to be in complete accordance with the derivative discontinuity of $E[N,N_s,v,B]$ with respect to $N_s$. Corresponding results in the $(N_\uparrow, N_\downarrow)$ representation can be similarly obtained, with derivatives with respect to $N_\uparrow$ and $N_\downarrow$ replacing the derivatives with respect to $N$ and $N_s$, and $B(\vec{r})$'s nonuniqueness requiring a derivative discontinuity of $F^L_{N_\uparrow,N_\downarrow}[n_\uparrow,n_\downarrow]$ both in $N_\uparrow$ and $N_\downarrow$.

**Acknowledgments:** T.G. acknowledges a grant from the Fund for Scientific Research – Flanders (FWO).

## Appendix: $F^L_{N,N_s}[n,s]$ for $\int n(\vec{r})d\vec{r} \neq N$ and $\int s(\vec{r})d\vec{r} \neq N_s$

In this Appendix, we show why the original Lieb definition for the SDFT $F$ functional, $F^L_{N,N_s}[n,s]$, gives infinity for $\int n(\vec{r})d\vec{r} = \tilde{N} \neq N$, and for $\int s(\vec{r})d\vec{r} = \tilde{N}_s \neq N_s$. Consider a constant external potential, $v(\vec{r}) = \bar{v}$. For that, the expression the supremum of which is taken in $F^L_{N,N_s}[n,s]$'s definition [Eq.(9)] can be written as $E[N,N_s,0,B] + N\bar{v} - \tilde{N}\bar{v} + \int s(\vec{r})\beta_e B(\vec{r})d\vec{r}$. As $\bar{v}$ is increased (decreased) infinitely in the case of $N > \tilde{N}$ ($N < \tilde{N}$), the value of the expression tends to infinity. This means that the supremum is infinity. Now, consider a constant external magnetic field, $B(\vec{r}) = \bar{B}$. The expression the supremum of which has to be taken gives $E[N,N_s,v,0] - N_s\beta_e\bar{B} - \int n(\vec{r})v(\vec{r})d\vec{r} + \tilde{N}_s\beta_e\bar{B}$ for $\bar{B}$, which similarly tends to infinity as $\bar{B}$ is decreased (increased) infinitely. This, again, yields an infinite supremum.